\RequirePackage{lineno}

\documentclass[3p,times]{elsarticle}
\usepackage{ecrc}

\volume{00}

\firstpage{1}

\journalname{Nuclear Physics A}

\runauth{A. Ohlson}

\jid{nupha}

\usepackage{amssymb}

\usepackage[figuresright]{rotating}

\begin{document}

\begin{frontmatter}



\dochead{}

\title{Jets and Jet-like correlations in STAR}


\author{Alice Ohlson for the STAR Collaboration}

\address{Physics Department, Yale University, New Haven, CT, USA.}

\ead{alice.ohlson@yale.edu}

\begin{abstract}
The propagation and modification of hard-scattered partons in the QGP can be studied using various types of jet and jet-like correlation measurements. The STAR detector with its full azimuthal and large pseudorapidity acceptance, as well as its wide transverse momentum ($p_T$) coverage, is well-suited for these measurements. At mid-rapidity, azimuthal correlations of charged hadrons with the axis of a reconstructed trigger jet are used to study the modification of jet shapes and associated hadron yields from $p$+$p$ to Au+Au. Dihadron correlations with back-to-back high-$p_T$ hadron pairs are used to investigate dijets and fragmentation biases.  STAR's increased particle identification capabilities due to the Time-Of-Flight detector are utilized to investigate the differences between jet-related and bulk-related particle production. Dihadron correlations with identified trigger particles provide experimental tests of simple recombination theories. The comprehensive set of STAR jet-quenching measurements can be used to further constrain theories of parton energy loss at RHIC.
\end{abstract}


\end{frontmatter}


\section{Introduction}
Analyses of dihadron correlations produced some of the first evidence that a new state of matter was created in $\sqrt{s_{NN}} = 200$ GeV Au+Au collisions at the Relativistic Heavy Ion Collider (RHIC) \cite{STARdihadron}.  In these analyses, distributions of the relative azimuthal angle ($\Delta\phi$) and relative pseudorapidity ($\Delta\eta$) between a high-$p_T$ ``trigger'' hadron and ``associated'' charged particles in an event are constructed.  These correlations show a ``nearside'' peak around $(\Delta\phi,\Delta\eta) = (0,0)$ which represents the charged particles associated with the jet containing the trigger, and an ``awayside'' peak at $\Delta\phi = \pi$ which consists of the hadrons associated with the recoil jet.  The modification of the shapes and associated particle yields of these jet peaks between $p$+$p$, $d$+Au, and Au+Au can be studied as a function of the transverse momentum of the trigger particle ($p_T^{trig}$) and of the associated particles ($p_T^{assoc}$).  The disappearance of the awayside peak in Au+Au \cite{STARdihadron1}, observed in early RHIC measurements, was attributed to parton energy loss in the quark-gluon plasma (QGP)~\cite{radTheoryFirst,radTheory0}.  

The Solenoidal Tracker at RHIC (STAR) detector is well-suited to correlation analyses due to its full ($2\pi$) azimuthal acceptance at mid-rapidity ($|\eta| < 1$).  In this region, the Time Projection Chamber (TPC) \cite{STARtpc} is used for charged particle tracking and particle identification (PID), the Barrel Electromagnetic Calorimeter (BEMC) \cite{STARbemc} is used to determine the neutral energy component of events, and the Time-Of-Flight detector (TOF) \cite{STARtof} provides additional PID capabilities.  Pions, kaons, and protons with transverse momentum in the range $0.7 < p_T < 4.0$ GeV/$c$ can be identified using a combination of TPC and TOF information.  

The concept of dihadron correlations has been extended and modified in different ways in order to address multiple physics questions, several of which will be described here.  Although many correlation analyses have been performed at STAR, only high-$p_T$, triggered correlations at mid-rapidity will be discussed.  

\section{Jet-hadron Correlations -- extending the kinematic reach to probe partonic energy loss}
Advances in jet-finding techniques \cite{fastjet} make it possible to use reconstructed jets as triggers in correlation analyses.  This extends the kinematic reach of these analyses because jet reconstruction can sample a higher-energy parton population than single particle measurements.  

In the jet-hadron correlation analysis, trigger jets are reconstructed with the anti-$k_T$ algorithm \cite{antikt,fastjet2}.  An online high tower (HT) trigger (which requires a transverse energy $E_T > 5.4$ GeV to be deposited in a single tower of the BEMC) selects events which contain high-$p_T$ processes.  An offline software cut raises the HT trigger threshold to $E_T > 6$ GeV.  Only tracks with $p_T > 2$ GeV/$c$ and towers with $E_T > 2$ GeV are used in the jet reconstruction, and the jet must contain (have as one of its constituents) a BEMC tower which fired the HT trigger.  The HT trigger requirement and $p_T$ cut are used to bias the trigger jet population towards hard fragmentation, control the effects of background fluctuations, and make the comparision between Au+Au and $p$+$p$ more straightforward.  The highly-biased nearside jet is used to assign conservative uncertainties on the shape of the combinatoric background (modulated by $v_2$ and $v_3$) and the trigger jet energy scale.  

\begin{figure}[tb]
\begin{minipage}{0.5\linewidth}
\includegraphics[width=\linewidth]{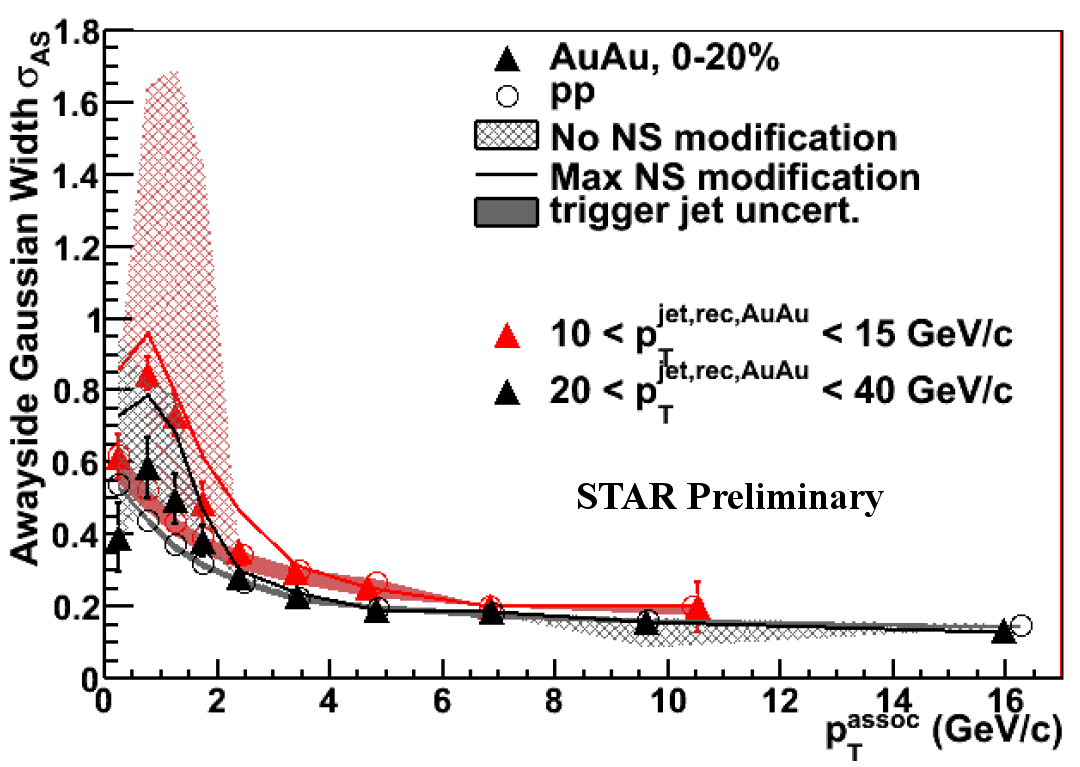}
\end{minipage}
\begin{minipage}{0.5\linewidth}
\includegraphics[width=\linewidth]{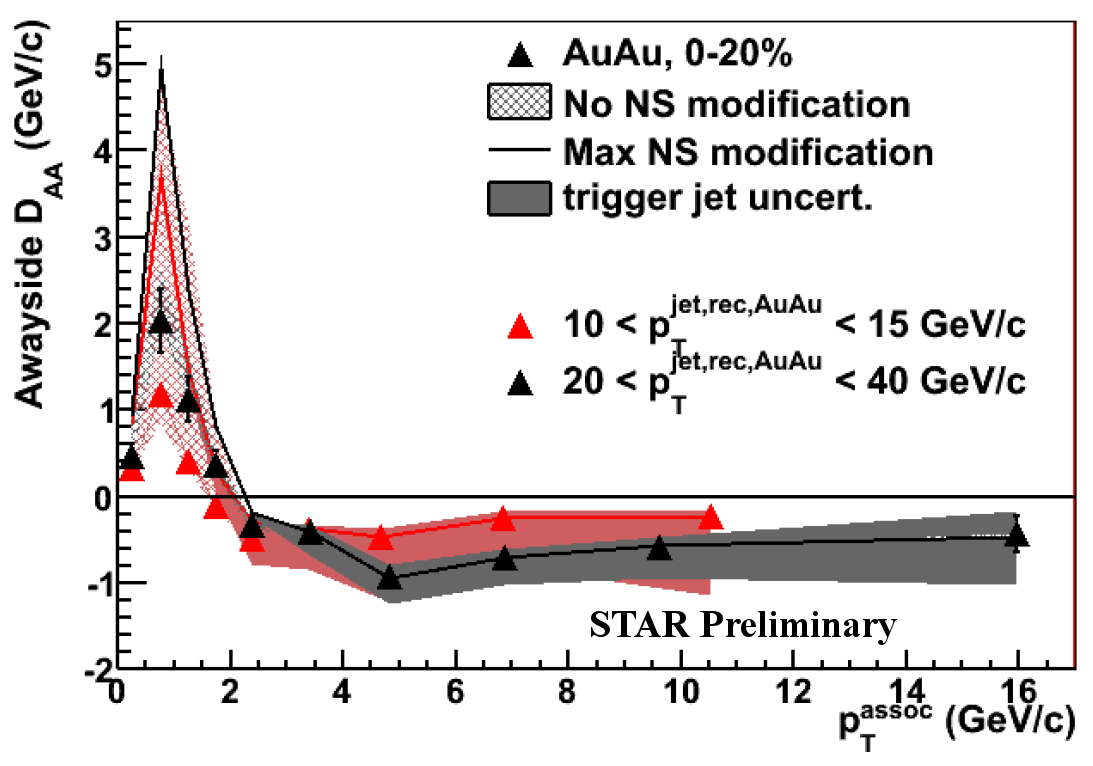}
\end{minipage}
\caption{\label{fig:jh}The awayside widths (left) and $D_{AA}$ (right) are measured in the jet-hadron correlation analysis, and are shown for two reconstructed jet $p_T$ ranges: 10 -- 15 GeV/$c$ and 20 -- 40 GeV/$c$.  The hatched bands are due to uncertainty in the shape of the heavy ion background ($v_2$ and $v_3$), while the shaded bands are due to uncertainties in the trigger jet energy scale.}
\end{figure}

In order to quantify the effects of jet-quenching on the awayside, the Gaussian widths of the awayside peak in $p$+$p$ and Au+Au are compared.  Additionally, the energy difference $D_{AA}$ (Equation~\ref{eq:daa}) and the energy balance $\Delta B$ (Equation~\ref{eq:deltab}) are calculated.  If jets in Au+Au fragment as in vacuum, then $D_{AA}$ and $\Delta B$ would be zero.  The results are shown in Figure~\ref{fig:jh} and Table~\ref{tab:deltab}.  
\begin{linenomath}\begin{equation}\label{eq:daa}
D_{AA}(p_T^{assoc}) \equiv Y_{Au+Au}(p_T^{assoc}) \cdot \langle p_T^{assoc}\rangle_{Au+Au} - Y_{p+p}(p_T^{assoc}) \cdot \langle p_T^{assoc}\rangle _{p+p}
\end{equation}\end{linenomath}
\begin{linenomath}\begin{equation}\label{eq:deltab}
\Delta B \equiv \sum_{p_T^{assoc}\mbox{ bins}} D_{AA}(p_T^{assoc})
\end{equation}\end{linenomath}

\begin{table}[tb]
\begin{minipage}{0.35\linewidth}
\renewcommand{\arraystretch}{1.3}
\begin{tabular}{|c|c|}
\hline
$p_T^{jet}$ (GeV/$c$) & AS $\Delta B$ (GeV/$c$)\\
\hline
10-15 & $0.0^{+0.2 +7.1 +0.9}_{-0.1 -0.3 -0.0}$\\\hline
20-40 & $0.6^{+0.2 +1.5 +3.3}_{-0.1 -0.0 -0.0}$\\
\hline
\end{tabular}
\end{minipage}
\begin{minipage}{0.65\linewidth}
\caption{\label{tab:deltab}Awayside (AS) $\Delta B$ values.  The first set of systematic uncertainties is due to detector effects (such as tracking efficiency, etc), the second set is due to $v_2$ and $v_3$ uncertainties, and the third set reflects the uncertainties on the jet energy scale.}
\end{minipage}
\end{table}

The awayside Gaussian widths are highly-dependent on the magnitude of the $v_3$ modulation.  More information is needed about jet $v_2$ and jet $v_3$, which are the correlations between reconstructed jets and the $2^{nd}$- and $3^{rd}$-order participant planes, before conclusions about medium-induced jet broadening can be drawn.  At low-$p_T^{assoc}$, the awayside $D_{AA}$ shows enhancement in the associated hadron yield in Au+Au compared to $p$+$p$, while at high-$p_T^{assoc}$ the awayside jet is suppressed in Au+Au, indicating that the recoil jet which traverses the medium is ``softened.''  The $\Delta B$ values are small compared to the reconstructed jet energy, indicating that the high-$p_T$ suppression is in large part balanced by the low-$p_T$ enhancement.  

\begin{figure}[tb]
\begin{minipage}{0.6\linewidth}
\includegraphics[width=\linewidth]{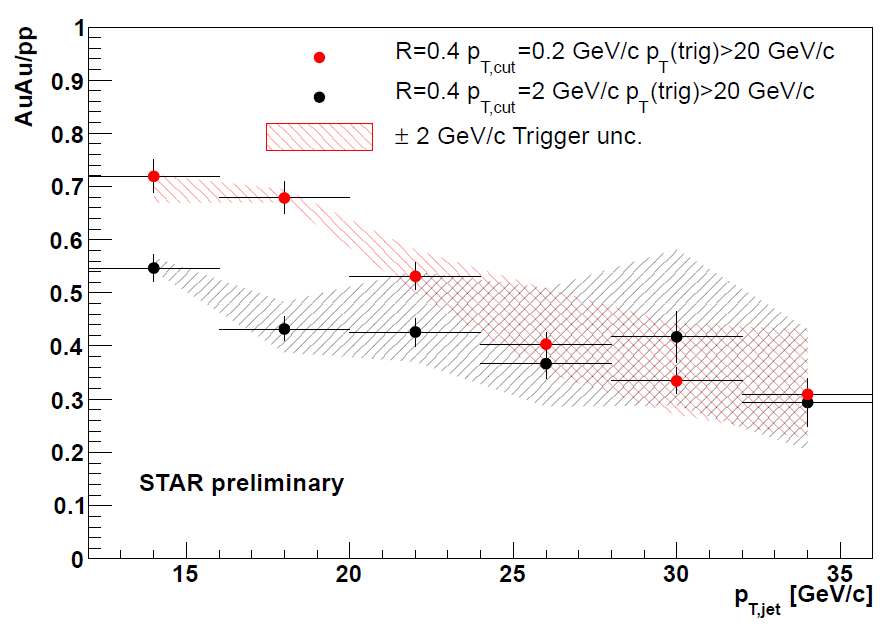}
\end{minipage}
\begin{minipage}{0.4\linewidth}\caption{\label{fig:dijet}  The ratio of the per-trigger recoil jet spectrum in Au+Au to the recoil jet spectrum in $p$+$p$ is shown~\cite{elenaHP}.  Trigger jets with $p_T^{trig} > 20$ GeV/$c$ are reconstructed with anti-$k_T$ ($R = 0.4$) using only TPC tracks with $p_T > 2$ GeV/$c$ and BEMC towers with $E_T > 2$ GeV.  The Au+Au/$p$+$p$ ratio is shown for recoil jets reconstructed with a $p_T$ cut of 2 GeV/$c$ (black) and 0.2 GeV/$c$ (red).  The recoil spectrum in Au+Au is significantly suppressed compared to $p$+$p$.}
\end{minipage}
\end{figure}

In addition to jet-hadron correlation measurements, the dijet coincidence rate in $p$+$p$ and Au+Au has also been measured~\cite{elenaHP}.  Figure~\ref{fig:dijet} shows the ratio between Au+Au and $p$+$p$ of the per-trigger $p_T$ spectrum of jets reconstructed on the recoil side of a reconstructed trigger jet with $p_T^{trig} > 20$ GeV/$c$.  The results are shown for recoil jets constructed with $R = 0.4$ and a constituent $p_T$ cut of 2 GeV/$c$ and 0.2 GeV/$c$.  In both cases the ratio is significantly below unity, indicating that the jet spectrum in Au+Au is suppressed compared to $p$+$p$, which may be due to softening of the recoil jet and/or broadening outside of the jet cone.  There is consistency between the conclusions of the jet-hadron correlation measurement and the dijet coincidence measurement.  

\section{2+1 Correlations -- exploiting fragmentation biases to probe pathlength dependence}
In the 2+1 correlation analysis, events are selected which contain a pair of back-to-back high-$p_T$ hadrons that serve as a dijet proxy.  Correlations can then be performed with respect to both trigger hadrons.  

The results for ``symmetric'' trigger pairs, in which the two trigger hadrons fall within similar $p_T$ ranges ($5 < p_T^{trig1} < 10$ GeV/$c$ and $4 < p_T^{trig2} < p_T^{trig1}$), were published in~\cite{2p1}.  It was shown that for this kinematic range of dijet triggers, and for associated hadrons within $1.5 < p_T^{assoc} < p_T^{trig1}$, there was no significant difference between the shapes of the nearside and awayside peaks, or between the correlations in Au+Au and $d$+Au.  These results indicate that the symmetric trigger requirement selects events with either surface-biased (``tangential'') or non-interacting dijets. 

\begin{figure}[tb]
\includegraphics[width=\linewidth]{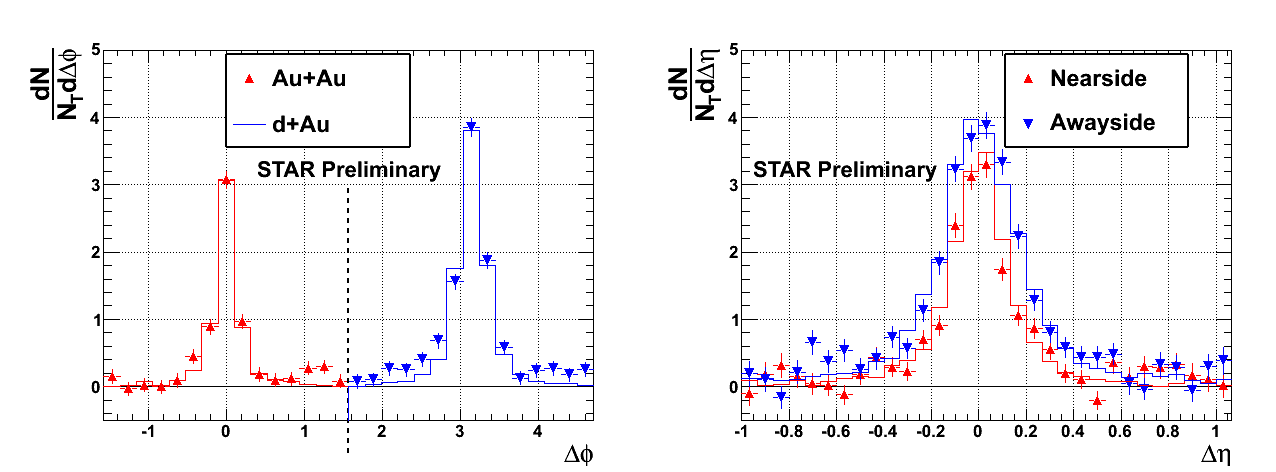}
\caption{\label{fig:2p1}The results of the 2+1 correlation analysis are shown for: $E_T^{trig1} > 10$ GeV, $p_T^{trig2} > 4$ GeV/$c$, and $p_T^{assoc} > 1.5$ GeV/$c$~\cite{huaWWND}.  (Left) The $\Delta\phi$ projections (within $|\Delta\eta| < 1$) show no significant difference in the jet peak shapes between Au+Au (solid triangles) and $d$+Au (solid lines).  (Right) The $\Delta\eta$ projections (within $|\Delta\phi | < 0.7$) show no significant difference between the nearside (red) and awayside (blue) peaks.}
\end{figure}

A similar analysis can be done with ``asymmetric'' trigger pairs~\cite{huaWWND}, in which the $p_T^{trig1}$ and $p_T^{trig2}$ thresholds are far apart.  The primary trigger is required to be a BEMC tower with $E_T^{trig1} > 10$ GeV, and the dijet trigger is a charged hadron with $p_T^{trig2} > 4$ GeV/$c$.  The correlations for this set of trigger kinematics are shown in Figure~\ref{fig:2p1} for $p_T^{assoc} > 1.5$ GeV/$c$.  There is still no significant peak shape difference between the nearside and the awayside, or between Au+Au and $d$+Au, in $\Delta\phi$ or $\Delta\eta$.  This is in contrast to the dihadron and jet-hadron correlation results, where modifications are seen.  

\begin{figure}[tb]
\includegraphics[width=\linewidth]{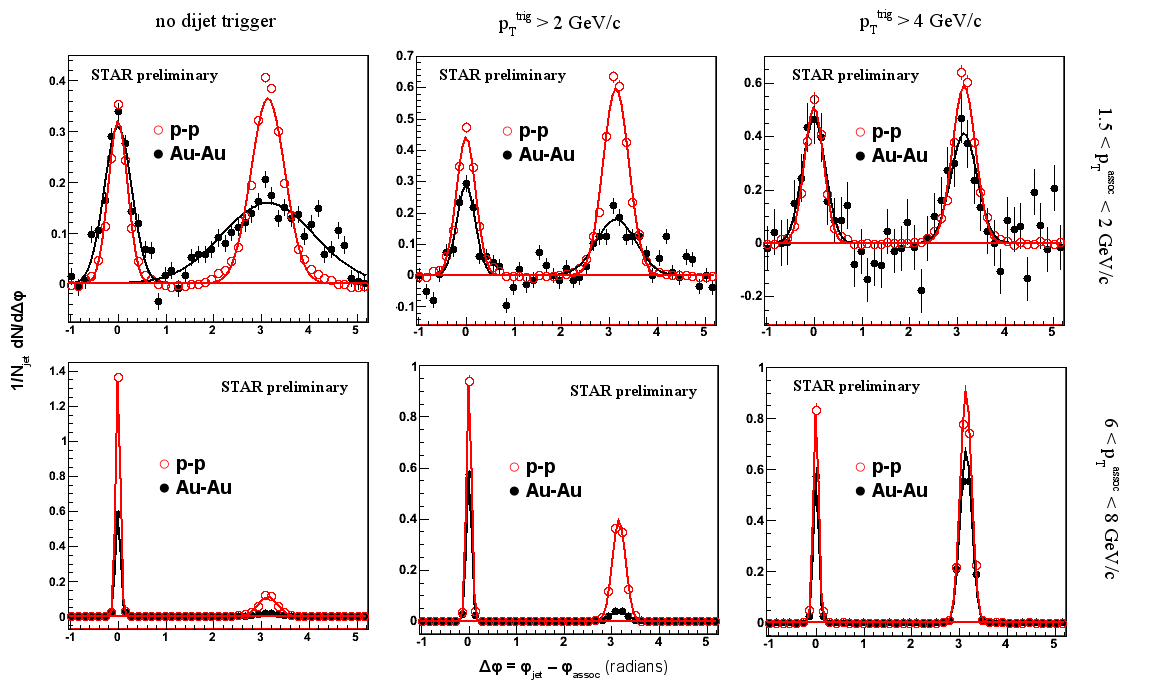}
\caption{\label{fig:2p1jh}Jet-hadron $\Delta\phi$ distributions are shown when requiring a charged hadron dijet trigger on the awayside ($|\phi_{jet} - \phi_{dijet\mbox{ }trig}| > \pi - 0.2$) for two $p_T^{assoc}$ bins: $1.5 < p_T^{assoc} < 2$ GeV/$c$ (top) and $6 < p_T^{assoc} < 8$ GeV/$c$ (bottom)~\cite{aliceWWND}.  The dijet trigger $p_T$ thresholds are: no dijet trigger (left), 2 GeV/$c$ (center), and 4 GeV/$c$ (right).  As the dijet trigger $p_T$ is increased, the awayside jet in Au+Au (closed black circles) looks more similar to $p$+$p$ (open red circles).}
\end{figure}

The biases involved in the 2+1 analysis can be explored by doing jet-hadron correlations with the additional requirement of a high-$p_T$ hadron on the awayside~\cite{aliceWWND}.  Figure~\ref{fig:2p1jh} shows the correlation functions as a function of the dijet trigger $p_T$ ($p_T^{trig2}$).  Suppression of the awayside jet peak in Au+Au is clearly seen when no dijet trigger is required (as well as possible broadening at low $p_T^{assoc}$, although $v_3$ has not been subtracted here).  When a 2 GeV/$c$ charged hadron is required opposite the trigger jet, the awayside jet peak in Au+Au becomes narrower, although it is still significantly suppressed compared to $p$+$p$.  When the dijet trigger threshold is raised to 4 GeV/$c$, as in the 2+1 analysis, the awayside jet peak in Au+Au appears very similar to $p$+$p$.  The 4 GeV/$c$ trigger hadron requirement largely selects unmodified jets which fragment as in vacuum.  

\section{PID in Correlations -- investigating jet and ridge composition}
Early RHIC measurements showed an enhanced baryon-to-meson (both $p/\pi^+$ and $\bar{p}/\pi^-$) ratio in central Au+Au collisions compared to peripheral Au+Au and $d$+Au~\cite{baryonmesonPHENIX,baryonmeson}.  This observation was successfully described by recombination models~\cite{recombination1,recombination2}.  A simplistic picture of recombination predicts that at intermediate $p_T$ there will be more baryons formed from thermal quarks than mesons.  Such a scenario would lead to a ``trigger dilution'' effect in dihadron correlation analyses when the trigger hadron is a baryon as opposed to a meson; there would be more baryon triggers without associated hadrons due to jet production, reducing the nearside per-trigger yield~\cite{triggerdilution}.  This hypothesis can be tested in a correlation analysis with identified trigger hadrons~\cite{koljaQM}.  

\begin{figure}[tb]
\begin{minipage}{0.5\linewidth}
\includegraphics[width=\linewidth]{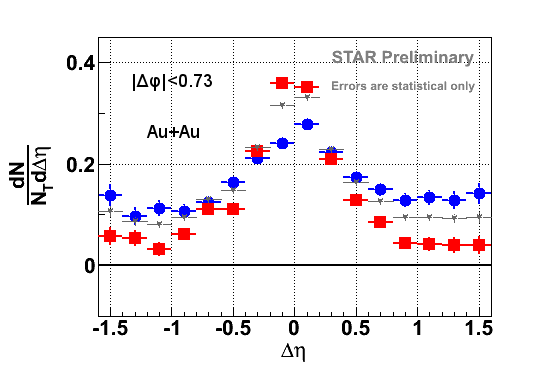}
\end{minipage}
\begin{minipage}{0.48\linewidth}
\includegraphics[width=\linewidth]{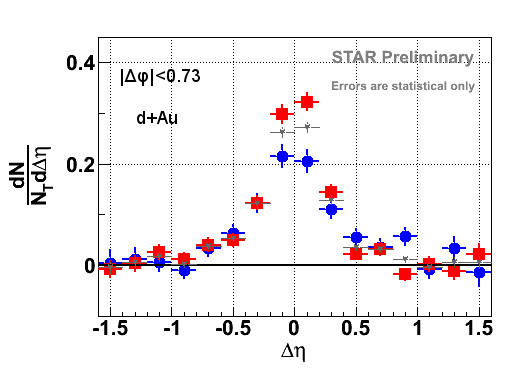}
\end{minipage}
\caption{\label{fig:PIDtrig}The nearside correlation structures are projected onto $\Delta\eta$ for $\pi^{\pm}$-triggered (red) and $(p^{\pm}+K^{\pm})$-triggered (blue) events in Au+Au (left) and $d$+Au(right)~\cite{koljaQM}.  It can be seen that in Au+Au the events with leading pions have a larger jet-like cone while the events with non-pion leading hadrons have a larger ridge/$v_3$ structure.  The difference in jet-like cone yield between pion- and non-pion-triggered events is also seen in $d$+Au.}
\end{figure}

Using the relativistic rise of the ionization energy loss ($dE/dx$) of charged particles in the TPC it is possible to statistically separate pions from kaons and protons.  The $\Delta\phi \times \Delta\eta$ correlation function can be split into two correlation functions, one in which the trigger hadron is a pion and another in which the trigger hadron is a kaon or proton.  The resulting correlation functions are projected onto $\Delta\eta$ (over $|\Delta\phi| < 0.73$) in order to examine the jet-like cone yield and ridge yield associated with different trigger species, and the results are shown in Figure~\ref{fig:PIDtrig}.  In Au+Au it is observed that there is a higher jet-like cone yield associated with $\pi^{\pm}$ triggers while the dominant contribution to the ridge or $v_3$ yield comes from the ($p^{\pm}+K^{\pm}$)-triggered events.  Although there is a difference in the jet-like cone yield for the pion- and non-pion-triggered events in Au+Au, that difference persists in $d$+Au, as seen in Table~\ref{tab:cone}.  The similarity between nearside yields in Au+Au and $d$+Au indicates that there is no trigger dilution, contrary to expectations from a simple recombination picture.  

\begin{table}[b]
\begin{minipage}{0.45\linewidth}
\renewcommand{\arraystretch}{1.3}
\begin{tabular}{|c|c|c|}
\hline
& Au+Au & $d$+Au \\
\hline
$\pi^{\pm}$ trigger & $0.22 \pm 0.01$ & $0.19 \pm 0.01$ \\\hline
$p^{\pm} + K^{\pm}$ trigger & $0.12 \pm 0.01$ & $0.14 \pm 0.02$ \\
\hline
\end{tabular}
\end{minipage}
\begin{minipage}{0.55\linewidth}
\caption{\label{tab:cone}The jet-like cone yield is shown for both trigger types in Au+Au and $d$+Au collisions.  Errors are statistical only.}
\end{minipage}
\end{table}

\begin{figure}[b]
\begin{minipage}{0.5\linewidth}
\includegraphics[width=\linewidth,trim=0mm 0mm 0mm 15mm,clip=true]{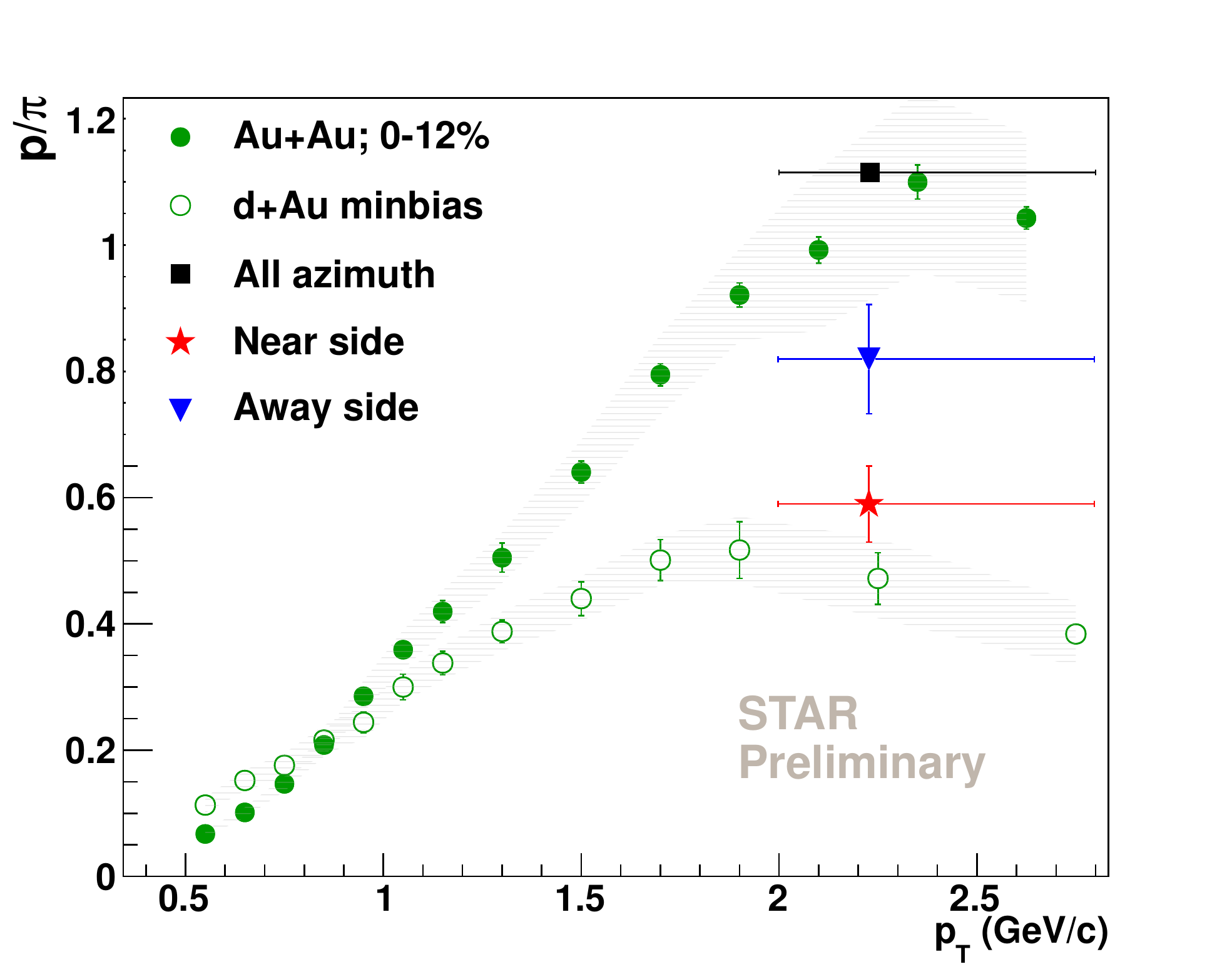}
\end{minipage}
\begin{minipage}{0.5\linewidth}\caption{\label{fig:ppi}The inclusive (all azimuth) $(p+\bar{p})/(\pi^++\pi^-)$ ratio is shown in events containing high-$p_T$ reconstructed jets, as well as the $(p+\bar{p})/(\pi^++\pi^-)$ ratio associated with the nearside (trigger) jet and awayside (recoil) jet \cite{alanHP}.  The results are compared to inclusive measurements in $d$+Au and central Au+Au collsions \cite{baryonmeson}. }
\end{minipage}
\end{figure}

The baryon/meson anomaly can be further investigated by performing correlation analyses with identified associated particles, to test whether the enhanced baryon production is correlated with jet or bulk processes.  In this jet-hadron correlation analysis, trigger jets are reconstructed with anti-$k_T$ ($R = 0.4$) from tracks with $p_T > 3$ GeV/$c$ and BEMC towers with $E_T > 3$ GeV.  The trigger jets are required to contain a HT trigger tower with $E_T > 5$ GeV.  Associated charged hadrons with $p_T^{assoc} < 2.8$ GeV/$c$ are identified using the TPC and TOF detectors.  The nearside and awayside proton and pion per-trigger yields are determined from fitting the $\Delta\phi$ distributions with two Gaussians on a constant background.  The ratios of these yields, shown in Figure~\ref{fig:ppi}, show a clear ordering in this kinematic range: the $(p+\bar{p})/(\pi^++\pi^-)$ ratio associated with the trigger jet is less than the $(p+\bar{p})/(\pi^++\pi^-)$ ratio associated with the awayside jet, which is less than the inclusive (all azimuth) ratio.  Due to the fragmentation biases induced by the HT trigger requirement and $p_T$ cut in jet-finding, comparisons to peripheral Au+Au and $p$+$p$ collisions are necessary before conclusions about baryon enhancement (or suppression) on the nearside are drawn.  For further discussion of this analysis, see~\cite{alanHP}.  

\section{Conclusions}
Correlation studies at RHIC have been utilized to explore the modification of jets that traverse the medium created in ultrarelativistic Au+Au collisions, and are compared to the cold nuclear matter baseline, $d$+Au, as well as to $p$+$p$ in which jets undergo vacuum fragmentation.  The correlation analysis technique has been extended to probe different physics questions.  

The kinematic reach of correlation analyses has been increased by using reconstructed jets as triggers, instead of high-$p_T$ hadrons serving as jet proxies.  The reconstructed trigger jet population is highly-biased towards hard fragmentation and thus likely consists of jets that undergo little medium-induced modification, potentially due to surface bias~\cite{Renk_Dihadron}.  This bias allows more straightforward comparison to trigger jets in $p$+$p$ collisions, and also increases the in-medium pathlength of the recoil parton, maximizing jet quenching effects on the awayside.  In the jet-hadron correlation analysis, the awayside jet peak is observed to be softened in Au+Au compared to the $p$+$p$ reference, which is qualitatively consistent with expectations from radiative energy loss models.  However, better experimental constraints on the shape of the underlying heavy ion background (in particular the value of jet $v_3$) are needed before conclusions about awayside broadening can be drawn.  A coincidence measurement of reconstructed recoil jets opposite a trigger jet also indicates that the awayside jets are suppressed due to softening and/or broadening outside of the jet cone in Au+Au.  

The 2+1 correlation analysis provides insight into the fragmentation biases induced by requiring a high-$p_T$ leading hadron.  An analysis of dihadron correlations in events containing a pair of back-to-back high-$p_T$ charged hadrons, where the kinematic cuts on the two trigger particles are similar, show no significant difference between shapes of the nearside and awayside jet peaks, or between Au+Au and $d$+Au.  The qualitative conclusions are similar even when the $p_T$ threshold of the leading hadron is increased dramatically, indicating that within the kinematic region chosen ($p_T^{trig1} > 5$ GeV/$c$ for the symmetric analysis and $E_T^{trig1} > 10$ GeV for the asymmetric case, $p_T^{trig2} > 4$ GeV/$c$, and $p_T^{assoc} > 1.5$ GeV/$c$), changing the $p_T^{trig1}$ cut does not select significantly different in-medium pathlengths, for either the nearside or awayside parton.  This can be further explored by performing the jet-hadron correlation analysis with a dijet trigger hadron required on the awayside.  It is observed that when the recoil jet contains a charged hadron with $p_T > 4$ GeV/$c$, the shapes of the recoil jet peaks in Au+Au and $p$+$p$ are similar, in contrast to the significant medium-induced modification observed when no such fragmentation bias is imposed on the awayside.  

Two analyses are discussed here which utilize the particle identification capabilities of STAR to probe the composition of the jet and the bulk.  It is seen that in dihadron correlations in which the leading particle is a pion, the jet-like cone yield is higher than in events where the leading hadron is a kaon or proton.  The ridge/$v_3$ component is dominated by events with non-pion triggers.  This difference in jet cone yield is also seen in $d$+Au events, which poses a challenge to simple recombination models that predict a trigger dilution effect in baryon-triggered correlations.  Jet-hadron correlations with identified associated particles show a difference between the $p/\pi$ ratio in jets when compared to the inclusive $p/\pi$ ratio, and will be used to further constrain models of jet-related and bulk-related particle production.

\bibliography{HPbiblio}

\begin{thebibliography}{10}
\expandafter\ifx\csname url\endcsname\relax
  \def\url#1{\texttt{#1}}\fi
\expandafter\ifx\csname urlprefix\endcsname\relax\def\urlprefix{URL }\fi
\expandafter\ifx\csname href\endcsname\relax
  \def\href#1#2{#2} \def\path#1{#1}\fi

\bibitem{STARdihadron}
C.~Adler, et~al., Phys. Rev. Lett. 90 (2003) 082302.

\bibitem{STARdihadron1}
J.~Adams, et~al., Phys. Rev. Lett. 91 (2003) 072304.

\bibitem{radTheoryFirst}
M.~Gyulassy, M.~Plumer, Phys. Lett. B243 (1990) 432--438.

\bibitem{radTheory0}
X.-N. Wang, M.~Gyulassy, Phys. Rev. Lett. 68 (1992) 1480--1483.

\bibitem{STARtpc}
M.~Anderson, et~al., Nucl. Instrum. Meth. A499 (2003) 659--678.

\bibitem{STARbemc}
M.~Beddo, et~al., Nucl. Instrum. Meth. A499 (2003) 725--739.

\bibitem{STARtof}
M.~Shao, O.~Y. Barannikova, X.~Dong, Y.~Fisyak, L.~Ruan, et~al., Nucl. Instrum.
  Meth. A558 (2006) 419--429.

\bibitem{fastjet}
M.~Cacciari, G.~P. Salam, Phys. Lett. B641 (2006) 57--61.

\bibitem{antikt}
M.~Cacciari, G.~P. Salam, G.~Soyez, JHEP 0804 (2008) 063.

\bibitem{fastjet2}
M.~Cacciari, G.~P. Salam, G.~Soyez, JHEP 04 (2008) 005.

\bibitem{elenaHP}
E.~Bruna, Nucl. Phys. A855 (2011) 367--370.

\bibitem{2p1}
H.~Agakishiev, et~al., Phys. Rev. C83 (2011) 061901.

\bibitem{huaWWND}
H.~Pei, J. Phys. Conf. Ser. 316 (2011) 012016.

\bibitem{aliceWWND}
A.~Ohlson, J. Phys. Conf. Ser. 316 (2011) 012015.

\bibitem{baryonmesonPHENIX}
S.~Adler, et~al., Phys.Rev. C69 (2004) 034909.

\bibitem{baryonmeson}
B.~Abelev, et~al., Phys. Rev. Lett. 97 (2006) 152301.

\bibitem{recombination1}
R.~C. Hwa, C.~Yang, Phys. Rev. C70 (2004) 024905.

\bibitem{recombination2}
R.~Fries, B.~Muller, C.~Nonaka, S.~Bass, Phys. Rev. C68 (2003) 044902.

\bibitem{triggerdilution}
R.~J. Fries, Nucl. Phys. A783 (2007) 125--132.

\bibitem{koljaQM}
K.~Kauder, J. Phys. G38 (2011) 124154.

\bibitem{alanHP}
A.~D\'{a}vila, these proceedings.

\bibitem{Renk_Dihadron}
T.~Renk, K.~Eskola, Phys. Rev. C75 (2007) 054910.

\end{thebibliography}
\bibliographystyle{elsarticle-num}


\end{document}